# Optimization of BaF$_2$ PAL-spectrometer geometry on basis of the Geant4 simulations

L.Yu. Dubov[2], V.I. Grafutin[1], Yu.V. Funtikov[1], Yu.V. Shtotsky[2],

L.V. Elnikova[1]

*[1] FSBI RF SSC Institute for Theoretical and Experimental Physics,*

*Bolshaya Cheremushkinskaya, 25, Moscow 117218, Russian Federation*

*[2]National Research Nuclear University «MEPhI»,*

*Kashirskoe shosse, 31, Moscow 115409, Russian Federation*

*E-mails:* Dubovl@mail.ru, grafutin@itep.ru, yshtotsky@mail.ru, elnikova@itep.ru

Incorrect choice of measuring geometry and signal discrimination thresholds for the BaF2 PAL spectrometer can lead to a significant distortion in the measured spectrum of positron lifetime. The contribution of the distorted events for non-optimal geometry (*e.g.* "face to face") can exceed 50%. It reduces the lifetime of the spectral components and redistributes intensities. Geant4 simulation permits to evaluate influence of measuring geometry on spectrum distortion and to choose the most appropriate discrimination thresholds. Optimal geometry with a lead absorber between the detectors permits to reduce contribution of distorted coincidences to less than 1% and to provide sufficiently high count rate of true events.

## 1. Introduction

Positron annihilation spectroscopy is a traditional method of non-destructive testing, allowing to study electronic structure of materials, mechanical and radiation-induced point and volume defects [1-2]. In measurements of the positron lifetime spectra, $^{22}$Na isotope, emitting simultaneously a positron and a nuclear photon with the energy 1.275 MeV, is most commonly used.

Positron lifetime is measured as a time delay of the annihilation photon signal relative to the *nuclear photon* signal. To determine the positrons lifetime (PLT) from measured spectra correctly, the accurate shape of spectrometer instrument function is required [3-5]. Measuring geometry and amplitude discrimination thresholds have a significant impact on the shape of instrument function and, therefore, on the shape of measured spectra and on reconstructed distribution of positron lifetimes.

To choose optimal measuring geometry and discrimination windows, that provide minimal distortion of the spectrum, and acceptable rate of the A-type coincidences, we have analyzed distortions of the instrument function on the basis of Monte Carlo simulation.

## 2. Simulation method

To develop Monte Carlo simulation of positron lifetime spectrometer, we applied Gate programming environment [5], which allows to describe detector systems by means of Geant4 libraries [6]. The model includes two cylindrical $BaF_2$ scintillators with a diameter of 25.7 mm and a height of 12.9 mm as detectors. The detectors may be placed at different distances ($L_1$ and $L_2$) from the center of the spectrometer opposite each other or be rotated at an angle $\varphi$. A plane isotropic source with a diameter of 5 mm is placed between them at a height $h$ from the horizontal axis. Besides the geometrical elements indicated in Fig. 1 the model also included aluminum shells of the detectors.

## 3. Analysis of distribution of delayed coincidences

To analyze the shape of delayed coincidence peak for $^{22}$Na source, we consider three successively emitted γ-rays: one nuclear photon (1.28 MeV) and two annihilation photons (0.511 MeV each). If only one and two-vertex physical events in the scintillators are taken into account there are ~ 65 possible event combinations which can produce coincidence. All combinations of coincidences may be divided into four types (see Fig. 2): "true" delayed coincidences (A-type), "self-coincidences" (B-type), "shifted" coincidences (C-type) and "mixed" coincidences (D-type).

The coincidences of the A-type are "true" coincidences of signals from a nuclear photon(1.28 MeV) in Start-detector and one of the annihilation photons, delayed by time interval $\tau$, in Stop-detector. The only reason of the broadening of A-type coincidences peak is the difference between distances traveled by photons to the points of interaction. The mean delay for A-type coincidences depends only on the positron lifetime $\tau$.

Coincident signals in both detectors can be also produced either by the single nuclear photon only or by the pair of annihilation photons only. Such coincidences are attributed to "self-coincidence" (B-type). For this type the delay distribution consists of three peaks. The peak in zero channel is a result of registering pair of annihilation photons by both detectors. Two shifted peaks are attributed to the case when the single nuclear photon is registered in both detectors. These peaks are shifted relative to the zero channel by the time $\pm\Delta t$, which depends upon the distance between the detectors. A delay for B-type coincidences does not depend on the positron lifetime $\tau$.

The C-type coincidences occur when both nuclear and one of annihilation photons delayed by $\tau$ are registered in the same either Start- or Stop-detector. It leads to a later triggering in the Start-channel or to an earlier triggering in the Stop-channel. The response time $t^*$ of a channel was defined as energy weighted mean: $t^* = (t_n \cdot E_n + t_a \cdot E_a) / (E_n + E_a)$, where $t_n$ and $t_a$ are registration moments for nuclear and annihilation photons respectively, while $E_n$ and $E_a$ are the energies absorbed in the scintillator. Thus, the time delay for C-type coincidences can be varied from 0 to $\tau$.

Coincidences of D-type are caused by double or triple events with time shifts $\Delta t$, $\tau$, or $\tau+\Delta t$. As a result, the delays for D-type coincidences may vary in a range from $-\tau$ to $\tau+\Delta t$.

This algorithm can be applied to choose the best measuring geometry and optimal values for discrimination windows which provide minimal distortion of positron lifetime spectrum, i.e.

minimize contribution of "B-D" types of coincidences, while retaining an acceptable efficiency (detection level) for A-type coincidences.

For simulation of lifetime spectra for different measuring geometries the following algorithm was used:

- Samples of coincidence events for nuclear and annihilation photons were obtained using the Geant4 simulation;

- Total energies and weighted mean coordinates for the events were calculated for every detector and energy-based selection was conducted;

- Stop signal delay was calculated for every coincidence event, on the basis of $\tau$, taken from the given distribution of positron lifetime;

- Delay times were distributed in accordance with additional broadening by spectrometer's instrumental function (normal distribution with a FWHM = 275 ps);

- A lifetime spectrum was built for a delay sample taking into account the width of the spectrum analyzer's channel (5 ps), and then was further processed using Palsfit software.

### 3. Results and discussion

Simulation was carried out for five different geometries (see Table 1). Geometry #1 (the detectors are placed at 10 mm from the center [$L_1 = L_2 = 10$ mm], thickness of the lead shield is 10 mm, the source is shifted by $h = 20$ mm relative to the central axis) is the optimal configuration. Geometry #5 commonly referred to as "face to face" (the closest distance between the detectors and the "central" position of the source) is traditional.

Every geometry under our study showed differences in resolution function defined by various contributions of A–D coincidence groups.

Fig. 3 shows the delay distributions for A-, B-, C-, and D-types of coincidences and the total peak ($\Sigma=A+B+C+D$) for the case when lifetime of each positron is $\tau \equiv 100$ ps. For the A-type coincidences a symmetric Gauss-like distribution with the $FWHM_A \approx 50$ ps is observed. Width of this distribution does not depend on the lifetime of the positron. The distribution of "self-coincidences" of the B-type consists of three peaks: a central peak and two symmetrically shifted ($\Delta t = \pm 40$ ps) lateral peaks with $FWHM_B \approx 30$ ps. Delay distribution of the C-type is non-symmetrical with a peak at about 75 ps with $FWHM_C \approx 60$ ps and its left side extends to zero channel. Time distribution of the D-type is even greater biased and extends from $-\tau$ to $+\tau$. The total spectrum for all coincidences, as shown in Fig. 3a, is essentially non-symmetrical and stretched significantly in the direction of lower lifetime.

Fig. 4 shows a shape of the total spectra for various lower thresholds of Stop-channel from 0.15 MeV to 0.45 MeV. Distortion of spectrum at low threshold increases dramatically.

Fig. 5 demonstrates shapes of total spectra for various fixed value of $\tau$ from 0 to 600 ps. When positron lifetime increases, the peak position is shifted in accordance with the value of $\tau$, while, the left part of the peaks is stretched up to $-\Delta t$. The shape of resolution function in this geometry depends significantly on positron lifetime $\tau$.

Fig. 6 shows the amplitude diagrams for A-, B-, C- and D-types of coincidences for non-shifted ($h = 0$ mm) and shifted ($h = 20$ mm) positions of positron source.

The diagrams show that "self-coincidences" (B-type) are completely excluded, if the sum of Start- and Stop-channels' lower thresholds exceeds 1.28 MeV. The "distorted" (C- and D-type) coincidences survive under any discrimination thresholds, but their number can be significantly reduced by displacement of the source. The number of "true" (A-type) coincidences is determined by the width and position of the discrimination windows.

The results of simulation (Table 1) show that count rate for the proposed measuring geometry #1 is much less than for the traditional "face to face" geometry (the number of "true" A-type coincidences is about 20 times less). However the contribution of "true" coincidences for geometry #1 is close to 100% versus some 60% for traditional geometry.

The results for simulated one-component spectra (0.1 ns, 0.4 ns and 2.0 ns) are presented in Fig. 7. We can see that only geometry #1 allows to recover the lifetimes with good accuracy (less than 1%). For other geometries, there is a significant "shortening" of the lifetime from 5 to 20%.

Fig. 8 shows the results for 2-component spectra: [0.1 ns (50%) + 0.4 ns (50%)] and [0.1 ns (50%) + 2.0 ns (50%)]. Lifetimes and intensities are recovered with good accuracy only for the optimum measuring geometry #1 (and partly for geometry #2). For other geometries a significant lifetime "shortening", 10% or more, is observed. In addition there is a substantial redistribution of component intensities: the intensity of the short-lived component increases from 50% to 60%, and the intensity of the long-lived component decreases, respectively. If the short-lived component (0.1 ns) is fixed, the long-lived component (0.4 ns or 2.0 ns) can be recovered with a good accuracy for all measuring geometries. However, redistribution of intensities in favor of a short-lived component remains.

Thus for any measuring geometries except for the optimal geometry #1 (and partly geometry #2), the distortion of component intensities always takes place: intensity of the short-lived component increases by 10–20 %, and intensity of the long-lived one decreases respectively.

## 5. Conclusion

For the traditional geometry $L_1=L_2=h=0$ we have the maximum count rate of the A-type ("true") coincidences (~ $10^3$ coincidences per $10^6$ decays). However, the fraction of the B–D-type coincidences, distorting the lifetime spectrum, is greater than ~ 35%. It makes practically impossible correct reconstruction of positron lifetime distribution.

In the measuring geometry with the detectors spaced 20 mm apart ($L_1 = L_2 = 10$ mm) and with a shifted source ($h = 20$ mm), the detection efficiency of the A-type ("true") coincidences is ~ $10^{-4}$, but the fraction of B–D-type coincidences decreases to 5–15%.

For the geometry with detectors rotated at 90º angle, with a lead shield arranged between the detectors, the fraction of the B–D-type coincidences is reduced to less than 2%. However, the detection effectiveness of the A-type coincidences is also about 3 times less.

The optimal geometry, *i.e.* shifted source and lead shield between the detectors (see Fig. 9), allows to reduce the contribution of distorting coincidences to an acceptable value.

For all measuring geometries, except the optimum one, there is a shortening of the evaluated lifetime components by 5-10% and a distortion of intensity ratio: the intensity of the short-lived component increases by 10-20%, while the intensity of the more long-lived component decreases respectively.

Lifetime spectra measured in a conventional "*face to face*" measuring geometry are significantly distorted, which leads to the "shortening" of evaluated lifetime components and to redistribution of their intensities in favor of the short-lived components.

The proposed measuring geometry, wherein the source-sample set is shifted by 20 mm from the axis of the detector system, the scintillation detectors are spaced 10 mm apart, and the lead screen in the form of a trapezoidal prism is positioned between detectors, allows to measure undistorted lifetime spectra and to keep an acceptable counting rate of events.

**Captions for figures and tables**

Fig. 1. The block diagram of a life-time spectrometer.

Fig. 2. Typical combinations of events for different types of coincidences from $^{22}$Na.

Fig. 3. The distribution of the delay time for $^{22}$Na, $\tau \sim 100$ ps, $L_1=L_2=h=0$, at the discrimination windows «Start»: 0.7–1.4 MeV and «Stop»: 0.35–0.6 MeV; (a) all the coincidences and the A-type «true» coincidences; (b) the B-, C-, and D-type coincidences.

Fig. 4. The spectra for the delayed coincidences ($L_1=L_2=h=0$) at the fixed $\tau \equiv 100$ ps for the lower threshold of the "Stop" channel discriminator varying from 0.1 to 0.4 MeV.

Fig. 5. The spectra of the delayed coincidences ($L_1=L_2=h=0$) at $\tau = 0$, 100, 200, and 600 ps, for the thresholds of the «Stop» (0.3–0.6 MeV) and «Start» (0.7–1.4 MeV) channels.

Fig. 6. The amplitude diagrams for the A-, B-, C-, and D-type coincidences ($^{22}$Na; «Start»: 0.6–1.5 МэВ; «Stop»: 0.1–0.6 МэВ; $L_1=L_2=0$) at $h=0$ (a) and $h=20$ mm (b).

Fig. 7. The results of the processing of one-component model spectra for five different geometries (see Table 1.).

Fig. 8. The refined results of the processing of modeled two-component spectra of 0.1 ns + 0.4 ns (A) and 0.1 ns + 2.0 ns (B) by Palsfit software with free parameters.

Fig. 9. The optimal measuring geometry.

Table 1. Contributions of the groups of coincidences into delayed coincidence spectrum and a number of "true" coincidences of the A-type (by $10^6$ positrons) for different measuring geometries at the thresholds of the differential discriminator "Start" 0.7÷1.4 MeV and "Stop" 0.35 ÷ 0.60 MeV.

**Figures**

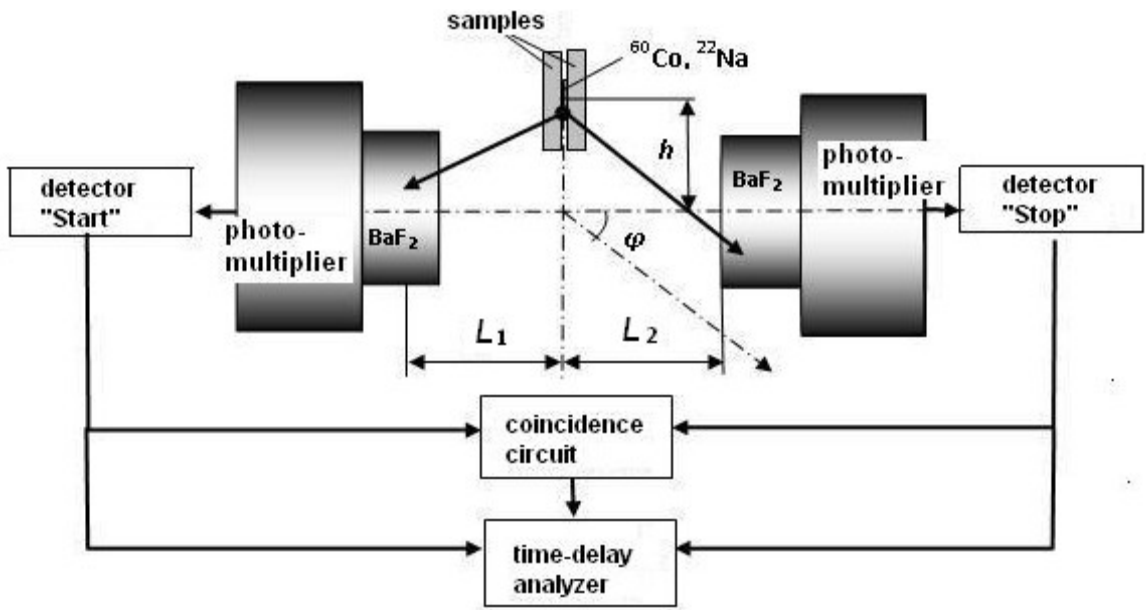

Fig. 1.

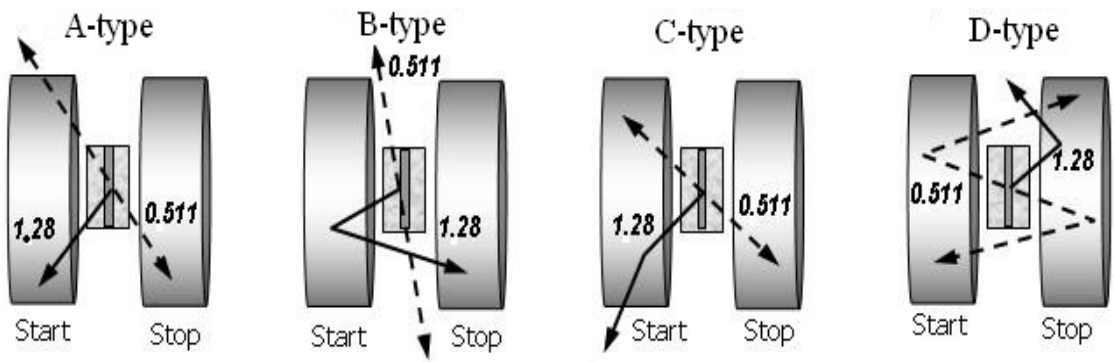

Fig. 2.

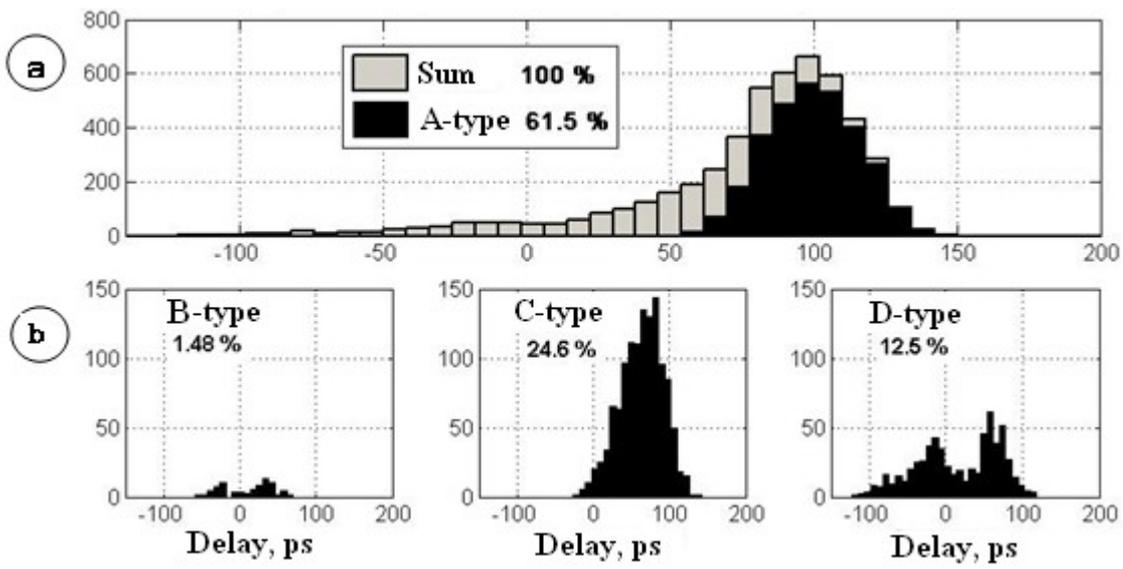

Fig. 3.

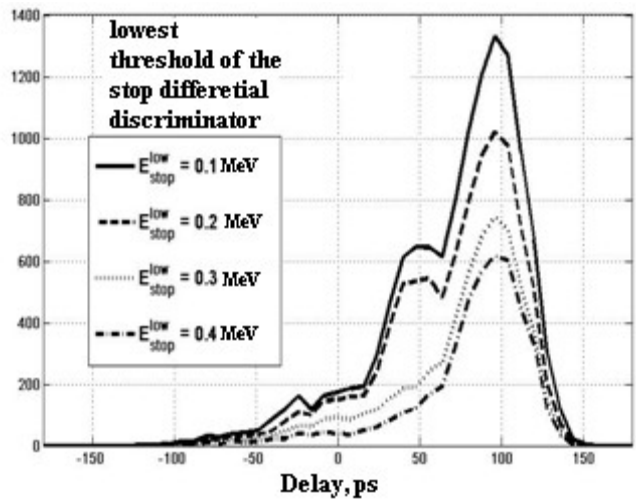

Fig. 4.

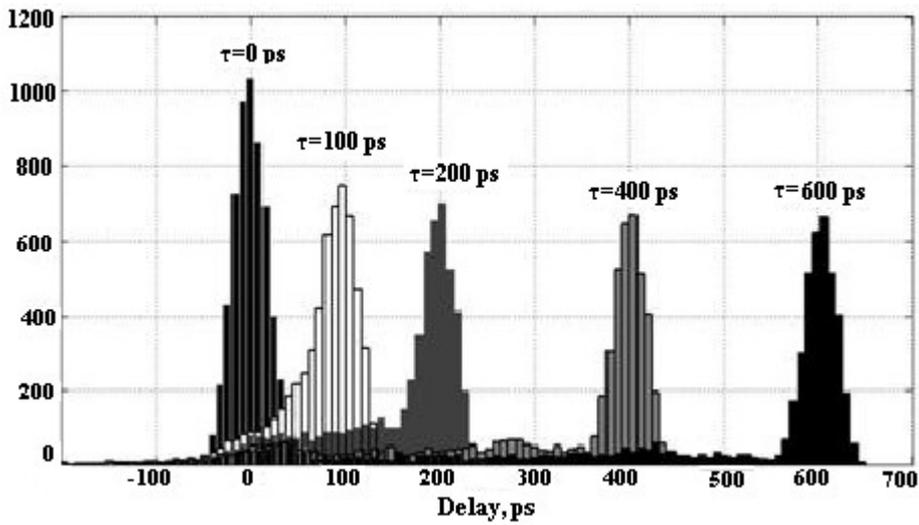

Fig. 5.

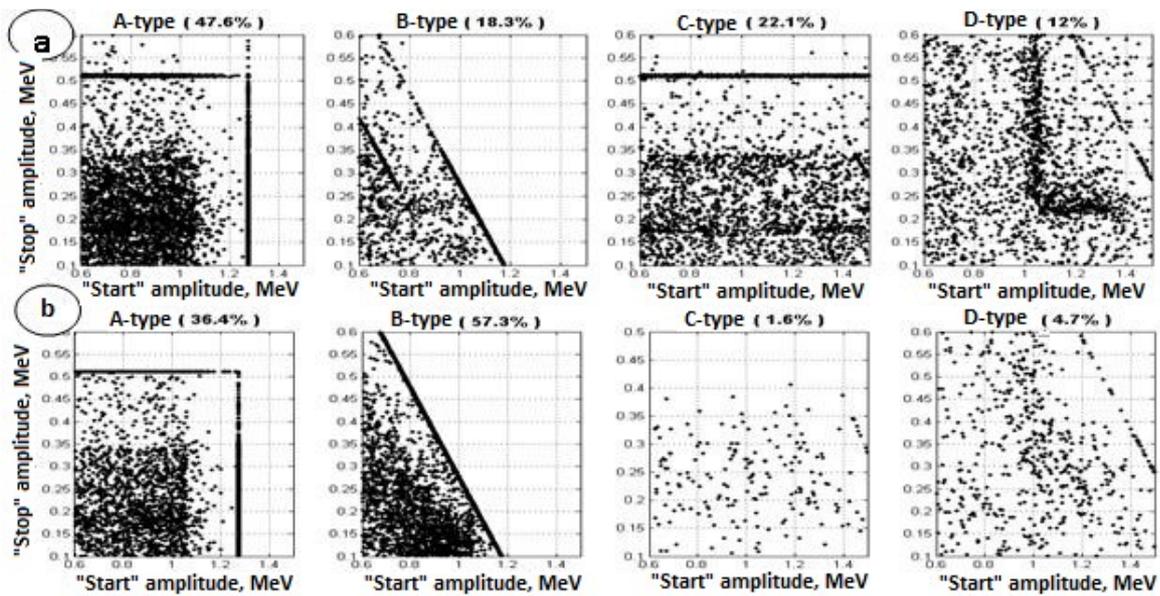

Fig. 6.

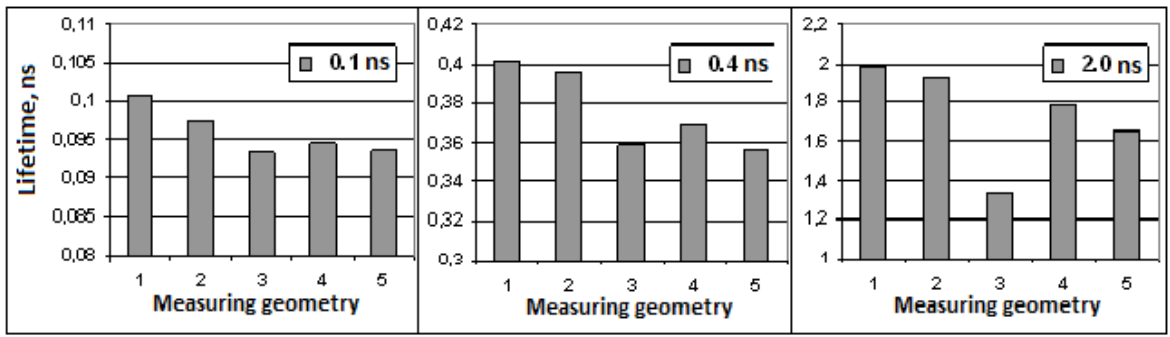

Fig. 7.

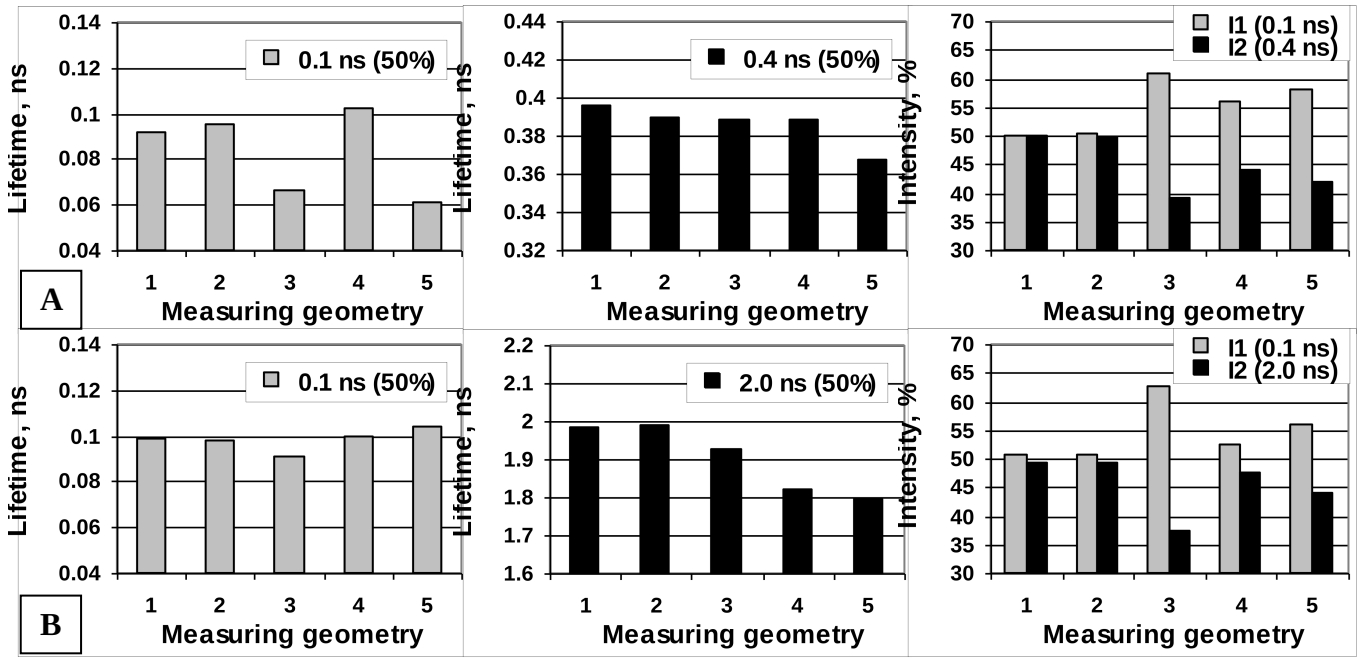

Fig. 8.

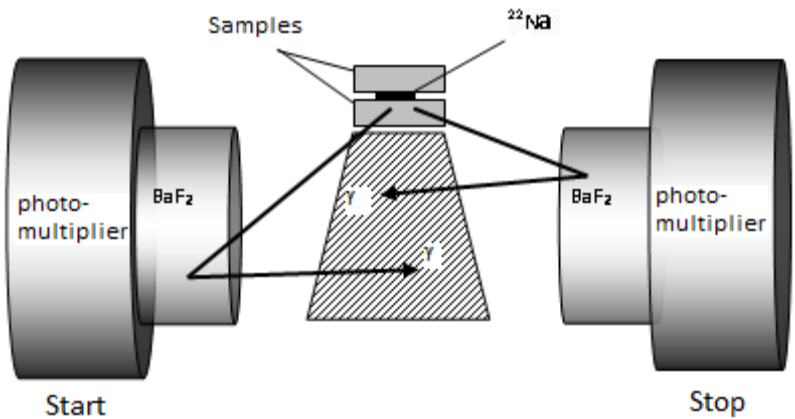

Fig. 9.

**Tables**

| Measuring geometry ($L_1$-$L_2$-$h$), mm | | | Contribution of coincidences, % | | | | $N_{«A»}$ |
|---|---|---|---|---|---|---|---|
| | | | A | B | C | D | |
| 110-10-20-Pb | #1 | Start [Pb] Stop | 99.5 | 0.3 | 0.2 | 0 | 115 |
| 10-10-20 | #2 | Start Stop | 96.4 | 1.1 | 0.3 | 2.2 | 241 |
| 0-0-20 | #3 | Start Stop | 70.4 | 20.8 | 0,5 | 8.3 | 344 |
| 10-10-0 | #4 | Start Stop | 72.7 | 0.5 | 23.9 | 2.9 | 345 |
| 0-0-0 | #5 | Start Stop | 61.5 | 1.5 | 24.5 | 12.5 | 3041 |